\documentclass[reprint,amsmath,amssym,aps]{revtex4-2}

\usepackage{graphicx}
\newcommand{\apjl}{Astrophys. J. Lett}
\newcommand{\mnras}{Mon. Not. Roy. Astron. Soc.}
\newcommand{\physrep}{Phys. Rep.}
\begin{document}

%\preprint{}

%\title{Circular orbits around dark-matter admixed neutron stars}
\title{Some properties of doubly-degenerate stars}

\author{Shin'ichirou Yoshida}
\email{yoshida@ea.c.u-tokyo.ac.jp}
\author{Junya Tanaka}
%\thanks{}
\affiliation{Department of Earth Science and Astronomy, Graduate School of Arts and Sciences, The University of Tokyo, \\Komaba 3-8-1, Meguro-ku, Tokyo 153-8902, Japan}


\date{\today}

\begin{abstract}
We investigate critical masses of and circular geodesics around 
doubly-degenerate stars (DDSs) which are composed
of cold nuclear matter as well as cold Fermionic dark matter (DM).
We here consider asymmetric dark Fermion with self-interaction as a DM candidate.
These stars have core-envelope structures and are categorized into baryon-enveloped 
and DM-enveloped, according to the composition of their envelope.
It is seen that the baryon-enveloped and DM-enveloped classes have their own critical
masses determined mainly by the dominant component in their envelope.
For a typical parameter sets, we see that a balanced mixture of two species may
lead to smaller masses than if the either of the species is dominant.
 We also show that for a highly DM-enveloped
case circular orbits in the vacuum region terminates at the innermost stable 
circular orbit (ISCO) in vacuum, but circular orbits of smaller radius are possible in the 
DM-envelope forming a gap between the ISCO and the inner orbits.
\end{abstract}

%\keywords{}

\maketitle

%%%%%%%%%%%%%%%%%%%%%%%%%%%%%%%%%%%%%%%%%%%%%%%%%
%%%%%%%%%%%%%%%%%%%%%%%%%%%%%%%%%%%%%%%%%%%%%%%%%
%%%
%%%   Section I : Introduction
%%%
%%%%%%%%%%%%%%%%%%%%%%%%%%%%%%%%%%%%%%%%%%%%%%%%%
%%%%%%%%%%%%%%%%%%%%%%%%%%%%%%%%%%%%%%%%%%%%%%%%%
%%%%%%%%%%%%%%%%%%%%
\section{Introduction}
%%%%%%%%%%%%%%%%%%%%
%%%
The nature of cosmic dark matter (DM) still remains one of the 
most perplexing mysteries of the physics for more than 80 years
since Zwicky suggested its presence in clusters of galaxies \cite{1933AcHPh...6..110Z,1937ApJ....86..217Z}.
The necessity of DM in galactic scale
\cite{1980ApJ...238..471R}, in clusters of galaxies (e.g., \cite{2006ApJ...648L.109C}),
and in cosmological scale \cite{2006Natur.440.1137S} is now firmly established. 
Baryonic astronomical objects that emit too weakly to be detected
are severely constrained as dark matter candidates from the Big Bang
nucleosynthesis. Most promising candidates are yet-to-be-found
elementary particles that do not interact with ordinary matter or at least do very weakly. \cite{2012AnP...524..507F}
From the point of view of the formation of cosmic structures
the dark matter must be subrelativistic and in the standard theory
of DM it is regarded as collisionless (collisionless cold dark matter, CCDM).
There are, however, some astronomical observations that are at odds with
CCDM models. Dwarf galaxies have flatter density profile at their center
than is expected from CCDM \cite{1994Natur.370..629M,1994ApJ...427L...1F}.
Moreover our Galaxy should have more subhaloes than are observed
as satellite galaxies \cite{1993MNRAS.264..201K, 1999ApJ...522...82K, 1999ApJ...524L..19M}.
Also "Too-big-to-fail" problem \cite{2011MNRAS.415L..40B} exists for
the subhaloes of our Galaxy to be explained by CCDM.
These suggest that DM may be collisional. One of the possible modification
of CCDM models is to introduce the self-interaction of DM particles  
\cite{2012MNRAS.423.3740V,2013MNRAS.430...81R,2013MNRAS.431L..20Z,2013MNRAS.430..105P}.
They may naturally solve the issues of CCDM above.
One of the categories that allow the solution is asymmetric 
DM (ADM, see \cite{2014PhR...537...91Z} for a review). In the early
Universe the baryon to anti-baryon number ratio might be asymmetric and
leads to the baryon dominant Universe as it is now. The same may
hold for DM particles and the rest of the pair-annihilated may be observed
as DM now. If ADM is also self-interacting it may aggregate to form stellar-sized 
objects. This possibility has been studied in exotic models of compact
stars.
%-- dark matter objects alternative to BH
One of the suggestions made is that these exotic objects serve as alternatives
to the observed black hole candidates. This idea is severely constrained for accreting objects in X-ray 
binaries \cite{2004ApJ...606.1112Y}. 

%-- dark star
Objects with typical mass of neutron stars and made from dark matter 
are also considered.\footnote{
It should be noted that if DM particles are not asymmetric as in popular
WIMPs (weakly interacting massive particles) or axions, they may annihilate each other
and heat a neutron star after they are captured by it. These possibilities have been
tested by studying the cooling history of neutron stars (see, e.g., \cite{PhysRevD.77.023006}).
}
They are termed as dark stars \cite{2015PhRvD..92f3526K,2017PhRvD..96b3005M}
and possibility of them to be alternatives to neutron stars is discussed.
On the other hand, a neutron star may capture DM particles and the DM accumulated
together with nuclear matter (baryon component) form a new type of compact star.
In \cite{PhysRevD.84.107301} Leung et al. study the structure of these stars (dark matter admixed 
neutron stars, or DANS. See \cite{2019ApJ...884....9L} for an alternative formation scenario of DANS). 
This is a baryon-dominated counter part of dark stars 
(see also \citep{PhysRevD.85.103528} for their radial stability). We are here interested 
in the general cases that encompass both dark stars and DANS, which we call as 
doubly-degenerate stars 
\footnote{These objects are called 'double degenerate stars' in \cite{2008ChPhL..25.2378L}.
We have slightly modified the term because it may be confused with binary stars whose components
are degenerate stars.
}.
%-- DANS

%== about this paper
In this paper we study characteristics of equilibria composed of 
baryonic matter and Fermionic dark matter in complete degeneracy. An equilibrium
star is categorized either to baryon-enveloped star or to DM-enveloped one.
The former has a core composed of baryon and DM which is covered with
an envelope composed of purely baryonic matter. The latter has instead
an envelope composed of DM. One of our
interests is the critical mass beyond which a stable equilibrium star cannot
exists. Although this has been studied in former studies, we need to pay
a careful attention to find the critical parameter when a star is composed
of multi-component fluid.
Another character of equilibria we study is the circular geodetic orbits of a test (baryonic)
particle. A neutron star in a X-ray binary have an accretion disk around it
whose radiation partly comes from the disk. The disk components of radiation depends
on the accretion state of the disk itself as well as the inner edge of the disk.
For a neutron star the inner edge is the smaller of either the surface of the star or 
the conventional innermost stable circular orbit (ISCO), which is located at the radius
of $6GM/c^2$ in Schwarzschild coordinate for a mass $M$ of the star. Unlike neutron stars with radii less than 
ISCO, we show for DDS with some parameter sets there appear multiple of stable orbits inside the ISCO.

Possible formation processes of these objects are not clear \cite{2015PhRvD..92f3526K,2019ApJ...884....9L}
and beyond the focus of our study here.
%%%%%%%%%%%%%%%%%%%%%%%%%%%%%%%%%%%%%%%%%%%%%%%%%
%%%%%%%%%%%%%%%%%%%%%%%%%%%%%%%%%%%%%%%%%%%%%%%%%
%%%
%%%   Section II: model
%%%
%%%%%%%%%%%%%%%%%%%%%%%%%%%%%%%%%%%%%%%%%%%%%%%%%
%%%%%%%%%%%%%%%%%%%%%%%%%%%%%%%%%%%%%%%%%%%%%%%%%
%%%%%%%%%%%%%%%%%%%%%
\section{\label{sec: formulation}Formulation}
%%%%%%%%%%%%%%%%%%%%%
\subsection{Assumptions}
We here study non-rotating stellar equilibria composed of baryonic and dark matter
whose mutual interaction is negligible. The baryonic matter is a zero-temperature nuclear
matter. The dark matter is a self-interacting Fermion with a vector
mediator field. 

\subsection{Tolman-Oppenheimer-Volkoff system for multi-component stars}
Since the stars in the present study are assumed to be static and spherically symmetric,
the spacetime allows the Schwarzschild coordinate $(r,\theta,\varphi)$
in which the spacetime metric is written as, 
\begin{equation}
	ds^2  = g_{\mu\nu}dx^\mu dx^\nu = -e^{2\nu}dt^2 + e^{2\lambda}dr^2 + r^2\sin^2\theta d\varphi^2.
\end{equation}
Here we assume the geometrized unit, e.g., $c=1=G$. The stress-energy tensor
of the matter is written as,
\begin{equation}
	T^{\mu\nu} = (\epsilon+p)u^\mu u^\nu + pg^{\mu\nu},\quad 
	\epsilon \equiv \epsilon_{_B}+\epsilon_{_X}, ~p \equiv p_{_B} + p_{_X}.
\end{equation}
The energy density $\epsilon$ and the pressure $p$ are simple summations 
of the baryonic component (with the subscript B)
and of the dark matter (with the subscript X). These thermodynamic variables obey
independent equations of state (EOS) which is described below. By using the Einstein's
equation and the extremal conditions of total mass with respect to the matter variables
by fixing the baryon number and the dark Fermion number, we obtain a modified 
Tolman-Oppenheimer-Volkoff system of equations used
in \cite{Ciarcelluti_Sandin2011} (see also \cite{Gresham_Zurek2019} and \cite{Kodama_Yamada1972}
for general cases with a finite interaction between baryonic and dark matter):
The equation for the metric potential $\nu$ is 
\begin{equation}
	\frac{d\nu}{dr} =  \frac{m+4\pi pr^3}{r(r-2m)},
	\label{eq: dnudr}
\end{equation}
where $m(r)=\int_0^r\epsilon 4\pi r'^2 dr'$.
The hydrostatic balance of particle species $i=B, X$,
\begin{equation}
	\frac{dp_i}{dr} =  -\frac{(\epsilon_i+p_i)(m+4\pi pr^3)}{r(r-2m)}.
	\label{eq: dpidr}
\end{equation}

%%%%%%%%%%%%%%%%
\subsection{Equation of state}
%%%%%%%%%%%%%%%%
\subsubsection{Dark matter}
As for the EOS of dark matter, we follow the treatment 
in \cite{Maselli2017} (see also \cite{Kouvaris2012}). The dark Fermions 
with the rest mass $m_{_X}$ are completely degenerated. They have a
self-interaction mediated by a boson with the rest mass $m_\phi$,
whose coupling constant is $\alpha_{_X}$. $\epsilon_{_X}$ and $p_{_X}$ are implicitly
related by 
\begin{eqnarray}
\epsilon_{_X} &=& \frac{m_{_X}^4}{\hbar^3}\left[\xi(x)+\frac{2}{9\pi^3}\frac{\alpha_{_X}}{\hbar}\frac{m_{_X}^2}{m_\phi^2}x^6\right] ;\nonumber\\
&&\xi(x) = \frac{1}{8\pi^2}\left[x\sqrt{1+x^2}(2x^2+1)-\ln(x+\sqrt{1+x^2})\right],
\label{eq: DMeos1}
\end{eqnarray}
and
\begin{eqnarray}
p_{_X} &=& \frac{m_{_X}^4}{\hbar^3}\left[\chi(x)+\frac{2}{9\pi^3}\frac{\alpha_{_X}}{\hbar}\frac{m_{_X}^2}{m_\phi^2}x^6\right] ;\nonumber\\
&&\chi(x) = \frac{1}{8\pi^2}\left[x\sqrt{1+x^2}\left(\frac{2}{3}x^2-1\right)+\ln(x+\sqrt{1+x^2})\right],
\label{eq: DMeos2}
\end{eqnarray}
In this paper we focus our interest on the repulsive self-interaction of dark Fermions, thus $\alpha_{_X}>0$.
\footnote{Attractive self-interaction does not necessarily mean a star is unstable to gravitational collapse
as shown in \cite{Gresham_Zurek2019}. We focus on the repulsive case for simplicity.
}
Here $x$ is the normalized Fermi momentum $x={\rm p}_{_X}/m_{_X}$. For our numerical computation
we solve for $v=x^2$. The value of $v=v_0$ at the origin parametrize the Fermi energy of DM.
It should be remarked that a solution of the so-called 'core-cusp' problem of galactic center may
require \citep{Maselli2017,2018PhR...730....1T}, 
\begin{equation}
	0.1({\rm g}{\rm cm}^{-3})\le 1.1\times \left(\frac{m_{_X}}{1{\rm GeV}}\right)\left(\frac{m_\phi}{10{\rm MeV}}\right)^{-4}\left(\frac{\alpha_{_X}}{10^{-3}}\right)^2 \le 10 ({\rm g}{\rm cm}^{-3}). 
\end{equation}
We use $m_{_X}=1$GeV, $m_\phi=10$MeV, $\alpha_{_X}=10^{-3}$
as a canonical set of parameters unless otherwise stated.

%%%%%%%%%%%%%%%%
\subsubsection{Baryonic matter}
%%%%%%%%%%%%%%%%
We adopt zero-temperature EOS for neutron star matter by \cite{Haensel_Potekhin2004, Potekhin_Chabrier2018}
and utilize the fortran subroutine {\tt nseos1.f} provided by
the authors at {\tt http://www.ioffe.ru/astro/NSG/NSEOS/}.
They made analytic fitting formulae for some representative nuclear EOS, 
e.g., FPS \citep{FPS1989}, SLy4\citep{SLy2001}, APR\citep{APR1998}. 
The stiffest of the three is APR, which produces the maximum mass
nonrotating neutron star with $M\sim 2.2M_\odot$, while the softest is FPS
by which a neutron star has a maximum mass of $M\sim 1.8M_\odot$
(see Fig.\ref{fig: MR relation}).

%%%%%%%%%%%%%%%%%%%%
\section{Results\label{sec: Results}}
%%%%%%%%%%%%%%%%%%%%

%============================================================
\subsection{Two kinds of equilibria: baryon-enveloped and dark matter-enveloped}
%================================
%%%%% Fig: R-M for different EOS %%%%
\begin{figure}[hptb]
\includegraphics[scale=0.6]{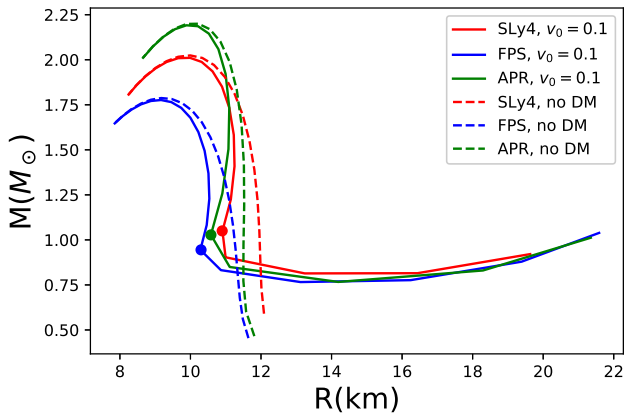}
\caption{Comparisons of mass-radius relation with/without dark Fermion. 
Dashed curves are purely baryonic neutron star sequences for various EOSs.
Solid curves are sequences that contain both baryons and dark Fermions.
Dark Fermion's momentum parameter is fixed as $v_0=0.1$. On each of
the DDS sequence the filled circle marks the boundary between baryon-enveloped
branch and DM-enveloped one. The branch with the larger radius corresponds
to the DM-enveloped.
\label{fig: MR relation}
}
\end{figure}
%%%%%%%%%%%%%%%%%%%%%%%
For the constant $v_0$ sequences, the inclusion of dark Fermion slightly modify 
the mass and the radius of the star from the purely baryonic counterpart
when the density is high enough. These are the branches in the left
of the filled circles on the solid curves in Fig.\ref{fig: MR relation}.
For these models, the radius of dark matter distribution $R_{DM}$
is smaller than the radius of the baryon distribution $R_B$, therefore
the stellar radius $R=R_B$. For $0\le r\le R_{DM}$ the baryonic and
dark matter coexist. We call these configuration as baryon-dominant.
On the righthand of the circles, the stellar configuration is mainly determined
by the degenerate pressure of the dark Fermion. The mass-radius relation
largely deviates from that of neutron stars. We have
$R_{DM}>R_B$ and the surface of the star $R=R_{DM}$. We call it
DM-dominant.

%================================
\subsection{Critical models of radial stability}
Now we consider the critical mass of the DDS beyond which
the star becomes unstable. For one parameter sequence of equilibrium
as cold neutron stars, the critical mass is the maximum mass
as a function of baryon density (or energy density).
With baryonic EOS and the dark Fermion EOS being fixed, 
the DDS models generally have two parameters that correspond to the energy
density of baryon and DM at the stellar center. In these multi-parameter
system, a (sufficient) stability criterion for family of equilibria is developed
by \cite{1982ApJ...257..847S} (see also \cite{1988ApJ...325..722F}). We follow their
treatment to find the critical mass of the DDS. 

We consider a sequence on which the baryonic mass $m_{_B}$ is kept constant. 
We may choose $\lambda\equiv v_0$ as a parameter to specify a model on the
sequence. And we assume an extremum of gravitational mass $M$
exists on the sequence at $\lambda=\lambda_0$. 
The theorem I of \cite{1982ApJ...257..847S}
tells us that if $\frac{d\mu_X}{d\lambda}\frac{dM}{d\lambda}>0$ on 
the secment of the sequence around $\lambda=\lambda_0$, it is unstable 
in the sense that an equilibrium is not a local minimum of $M$.\footnote{
In the original argument of \cite{1982ApJ...257..847S} 
a scalar $S$ to be maximized for an equilibrium
state is related to variables $E^a$ as $dS = \beta_a dE^a$.
As in \cite{1988ApJ...325..722F} we take $S = -M$ and an equilibrium
minimize it. Then we have $E^a=(-N_B, -N_X)$ and $\beta_a=(\mu_B, \mu_X)$.
Since an equilibrium star satisfies $dM = \mu_B dN_B + \mu_X dN_X$ we have
 $dM=\mu_X dN_X$ on the sequence with baryonic mass being fixed.
 Thus the stability condition reads $\frac{d\mu_X}{d\lambda}\frac{dM}{d\lambda}>0$}
Here $\mu_X$ is the chemical potential of dark Fermion including gravitational
contribution to energy, 
\begin{equation}
	\mu_{_X} = e^{\nu}\frac{\partial\epsilon_{_X}}{\partial n_{_X}},
\end{equation}
which is constant throughout an equilibrium star. 
%%%%% Fig: Mb-M for APR %%%%
\begin{figure}[hptb]
\includegraphics[scale=0.55]{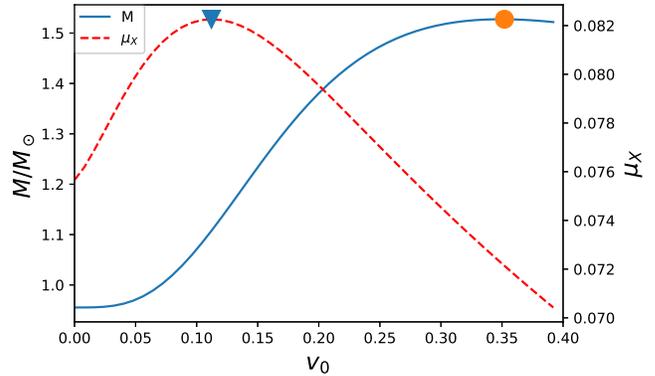}
\caption{Gravitational mass $M$ of an equilibrium sequence with baryonic
mass fixed as $M_{_B}=1.02M_\odot$. 
It is plotted as a function of $v_0$ parameter (solid). Limits
of $v_0\to 0$ are normal neutron star.
Also plotted (dashed) are dark Fermion's chemical potential $\mu_{_X}$
(in a arbitrary unit) for equilibrium star as a function of $v_0$.
The circle marks the maximum of $M$ and the triangle marks that of $\mu_{_X}$.
Baryonic EOS is APR and dark Fermion parameters are 
$(m_{_X}, m_\phi, \alpha_{_X})=(1{\rm GeV}, 10{\rm MeV}, 10^{-3})$. 
\label{fig: chemical potential and M}}
\end{figure}
%%%%%%%%%%%%%%%%%%%%%%%

Baryonic mass of a configuration is defined by the proper volume
integration of number density of the baryonic particle $n_{_B}$ as,
\begin{equation}
	M_{_B} = m_{_B}\int_0^R 4\pi r^2 e^{\lambda} n_{_B} dr,
	\label{eq: baryon mass} 
\end{equation}
where $m_{_B}$ is the mass of baryon particle (nucleon). In the same way the dark
matter mass is defined by the number density of dark Fermion $n_{_X}$
\begin{equation}
	M_{_{DM}} = m_{_X}\int_0^R 4\pi r^2 e^{\lambda} n_{_X} dr,
	\label{eq: DM mass} 
\end{equation}
where $m_{_X}$ is the mass of the dark Fermion.
Gravitational mass $M$ is defined as
\begin{equation}
	M = \int_0^R 4\pi r^2 \epsilon dr,
\end{equation}
where energy density is the sum of baryonic and dark Fermion contribution.

In Fig.\ref{fig: chemical potential and M} typical behavior of gravitational mass 
$M$ and
chemical potential $\mu_{_X}$ are shown on a sequence of $M_{_B}=$constant.
At the point $dM/d\lambda=0$ (marked by a circle), $\mu_{_X}$ has a negative slope. Therefore
models with larger $\lambda=v_0$ satisfy the condition of the instability
$\frac{d\mu_X}{d\lambda}\frac{dM}{d\lambda}>0$.
For this baryonic mass we conclude the critical model corresponds 
to this point. We remark that the models with smaller $v_0$ than that of the triangle
point also satisfies $\frac{d\mu_X}{d\lambda}\frac{dM}{d\lambda}>0$. It should be, however.
remembered that the stability criterion applies to the part of the sequence around the extremum
of $M$. Moreover the limit of $v_0\to 0$ corresponds to a pure neutron star model
with $M\sim 1.2M_\odot$. This limit is completely stable. We conclude this part of the sequence
is stable.

%%%%% Fig: Mb-M for APR %%%%
\begin{figure}
\includegraphics[scale=0.55]{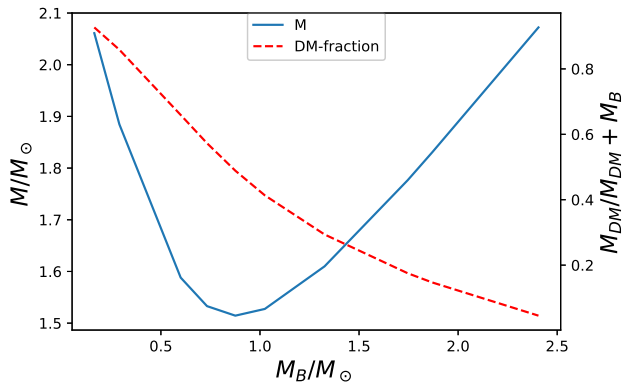}
\caption{(Solid) Maximum gravitational mass of DDS with APR as baryonic EOS.
(Dashed) Mass fraction of dark matter defined as $\frac{M_{_{DM}}}{M_{_B}+M_{_{DM}}}$.
The dark sector parameters are, $m_{_X}=1$GeV, $m_\phi=10$MeV, $\alpha_{_X}=1\times 10^{-3}$.
\label{fig: Mgmax-APR}
}
\end{figure}
%%%%%%%%%%%%%%%%%%%%%%%

In Fig.\ref{fig: Mgmax-APR} the critical gravitational mass (solid line) and dark matter fraction (dashed line)
defined by $M_{_{DM}}/(M_{_{DM}}+M_{_B})$ are plotted as functions of baryonic mass.
This is the case with APR EOS for baryons and dark Fermion parameters are
$(m_{_X}, m_\phi, \alpha_{_X})=(1{\rm GeV}, 10{\rm MeV}, 10^{-3})$.
It is remarkable that the critical mass is not a monotonic function. When $M_{_B}$ is large,
the DM fraction becomes small and the model is close to the pure neutron star. This is the
rightmost end of the plot. On the other hand for smaller and smaller $M_{_B}$ the star becomes
more and more DM-enveloped and the critical mass asymptotes to the pure DM star (dark
star limit). Between these limit there is a minimum of critical mass. For this particular model
parameter, nearly equal amount of contribution from baryon and DM does not support
a heavy star against its self-gravity.

%%%%%%%%%%%%%%%%%%%%%%%%%%%% Fig: Mb-M for FPS %%%%
\begin{figure}
\includegraphics[scale=0.55]{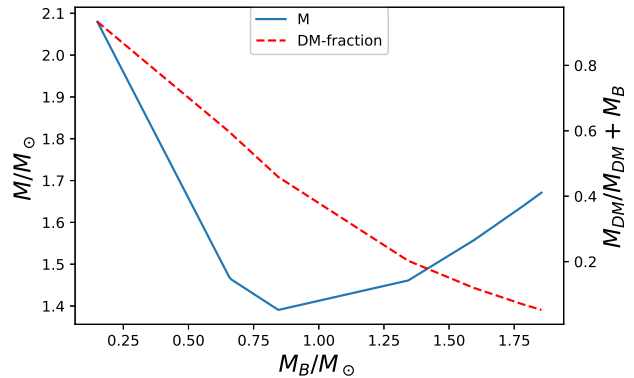}
\caption{Same as Fig.\ref{fig: Mgmax-APR} except that the baryonic EOS is FPS. The dark sector
parameters are the same.
\label{fig: Mgmax-FPS}
}
\end{figure}
%%%%%%%%%%%%%%%%%%%%%%%	
The similar characteristics holds for the softer baryonic EOS (Fig.\ref{fig: Mgmax-FPS})
although the mass supported by the pure baryonic matter is smaller.

%%%%% Fig: Mb-M for APR, mX=2GeV%%%%
\begin{figure}
\includegraphics[scale=0.55]{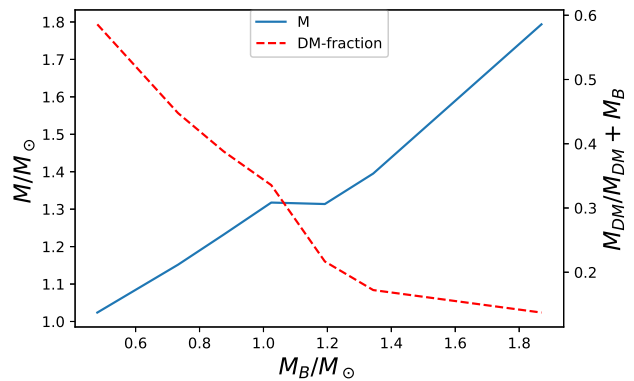}
\caption{Same as Fig.\ref{fig: Mgmax-APR} except the mass of the dark Fermion is $m_{_X}=2$GeV.
\label{fig: Mgmax-APR-mX2}
}
\end{figure}

When the DM parameters are modified, the characteristics may change.
In Fig.\ref{fig: Mgmax-APR-mX2} the mass and the DM fraction is plotted for APR EOS
and $(m_{_X}, m_\phi, \alpha_{_X})=2{\rm GeV}, 10{\rm MeV}, 10^{-3})$. This
case is compared with Fig.\ref{fig: Mgmax-APR} except that the dark Fermion mass is
larger. The critical
mass is now a monotonic function of $M_{_B}$. This is expected from considering the 
Chandrasekhar limit of a star made of Fermion of mass $m_0$ scales as
$M_{\rm Ch}\propto m_0^{-2}$. 

%%%%%%%%%%%%%%%%%%%%%%%%%%%% Fig: Mb-M for APR %%%%
\begin{figure}
\includegraphics[scale=0.55]{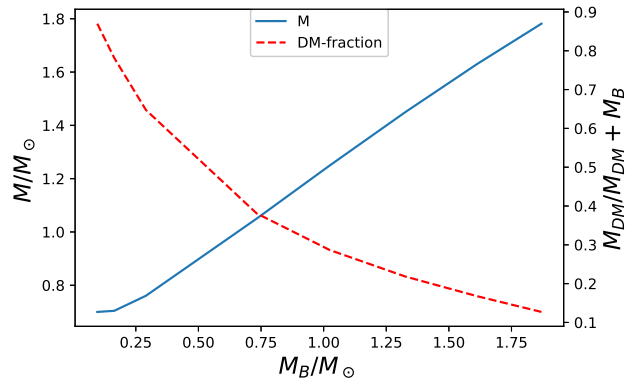}
\caption{Same as Fig.\ref{fig: Mgmax-APR} except that the mass of the mediator is $m_\phi=50$MeV.
\label{fig: Mgmax-APR-mphi5e-2}
}
\end{figure}
%%%%%%%%%%%%%%%%%%%%%%%
In Fig.\ref{fig: Mgmax-APR-mphi5e-2} we have the plots for the same parameters
except the mediator mass $m_\phi=50$MeV of the dark self-interaction
is larger than that of Fig.\ref{fig: Mgmax-APR}. In this case we also have small critical
mass for DM-enveloped limit. We expect it because the heavier mediator have smaller
Yukawa radius of interaction and it contribute less to the stiffness of the dark Fermion matter.

%%%%%%%%%%%%%%%%%%%%%%%%%%%% Fig: Mb-M for APR %%%%
\begin{figure}
\includegraphics[scale=0.55]{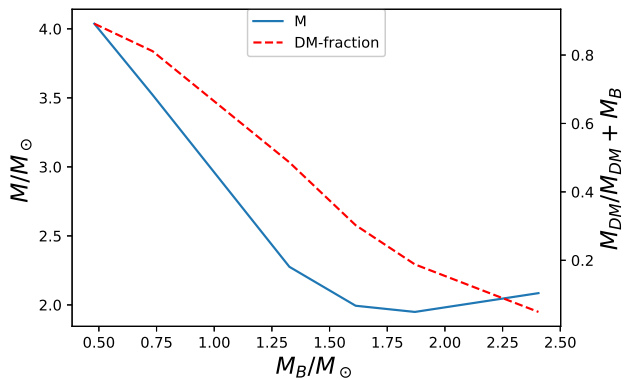}
\caption{Same as Fig.\ref{fig: Mgmax-APR} except that the coupling $\alpha_{_X}=5\times 10^{-3}$.
\label{fig: Mgmax-APR-aX5e-3}
}
\end{figure}
%%%%%%%%%%%%%%%%%%%%%%%
Finally Fig.\ref{fig: Mgmax-APR-aX5e-3} is the same plot as Fig.\ref{fig: Mgmax-APR} except that
$\alpha_{_X}$ is larger. This corresponds to the larger repulsive dark-interaction which results
in the larger mass of the DM-enveloped stars.

%===============================
\subsection{Circular orbits around/in DDS}
In this section we study the circular orbits of baryonic test particles around the DDS.
For a baryon-enveloped model we do not expect to see qualitative difference from
the normal neutron stars, that is, a circular orbit exists down to the surface of the star
if the stellar radius is larger than $6M$, or it is truncated at $6M$ if the radius is
smaller than $6M$. It is, however, not obvious if there exists stable circular
orbit inside the DM envelope of a DM-enveloped star, when the radius of the star
is smaller than $6M$. If circular orbits are allowed in the DM-envelope, accretion disks
(or rings) around the DDS may have qualitative difference from the standard
disk around the normal neutron stars. Therefore we test the possibility of circular
orbit in DM envelope of DM-enveloped DDS.
	
	Since the spacetime is spherically symmetric, we focus on the orbit with $\theta = \pi/2$
	without loss of generality. Let the 4-velocity of the particle to be $u^\mu=dx^\mu/d\tau$ 
	that satisfies the normalization $u^\mu u_\mu=-1$ (thus $\tau$ is the proper time). 
%	The world line of it extremizes the action,
%	\begin{equation}
%		I  = \int\left[-e^{2\nu}(u^t)^2 + e^{2\lambda}(u^r)^2 + r^2(u^\varphi)^2\right] d\tau.
%	\end{equation}
	We define two constants of motion, e.g. the mechanical energy $\epsilon$ and the specific
	angular momentum $\ell_0$,
	\begin{equation}
		\epsilon = -g_{t\beta}u^\beta = e^{2\nu}u^t,~\ell_0 = g_{\varphi\beta}u^\beta = r^2u^\varphi.
		\label{eq: specific e and l}
	\end{equation}
	Then the normalization condition of 4-velocity result in
	\begin{equation}
		E\equiv \frac{\epsilon^2-1}{2} = \frac{1}{2}\left(\frac{d\xi}{d\tau}\right)^2 + V_{\rm eff} ; 
		~ \xi \equiv \int e^{\nu+\lambda} dr,
		 \label{eq: energy-conserv}
	\end{equation}
	and the effective one-dimensional potential $V_{\rm eff}$ is defined as
	\begin{equation}
		V_{\rm eff} = \frac{e^{2\nu}}{2}\left(\frac{\ell_0^2}{r^2}+1\right) - \frac{1}{2}.
	\end{equation}
	Eq.(\ref{eq: energy-conserv}) may be seen as an equation for one-dimensional motion of a
	particle in a potential.  The left hand side may be regarded as a conserved total
	energy of the particle, while the first and the second terms on the right hand side are regarded as
	the kinetic and the potential energy $V_{\rm eff}$. Notice that $\epsilon$ includes
	the rest mass of test particle, thus it asymptotes to unity as $r\to\infty$.

%%%%%%%%%%%%%%%%% Veff %%%%
\begin{figure}[htbp]
\includegraphics[scale=0.55]{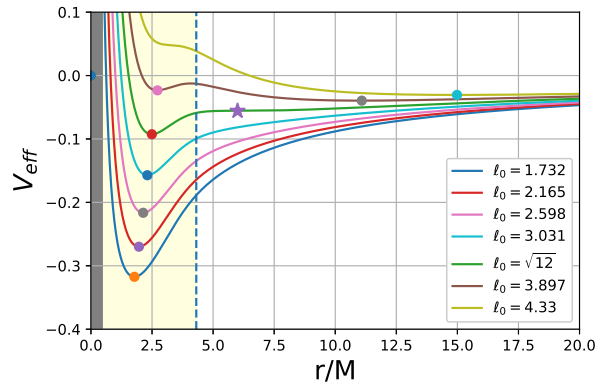}
\caption{Effective potential $V_{\rm eff}$ of reduced radial motion of a test particle for a star
with APR EOS and $(m_{_X}, m_\phi, \alpha_{_X})=(1{\rm GeV}, 10{\rm MeV}, 10^{-3})$.
The star is DM-enveloped with $\rho_c=2\times 10^{14}{\rm g}{\rm cm}^{-3}$ and $v_0=0.4$.
The gravitational mass $M=2.12M_\odot$, baryonic mass $m_{_B}=2.40\times 10^{-4}M_\odot$,
DM rest mass $M_{_{DM}}=2.25M_\odot$. The radius of the core of baryon and DM mixture
is $R_{in}/M=0.519$ and the radius of the DM envelope $R_{DM}/M=4.31$.
The dark-shaded area is the core and the light-shaded is the envelope whose outer edge
is highlighted by the vertical dashed line.
Each curve corresponds to the different value of specific angular momentum $\ell_0$
whose values are expressed in unit of $M$.
Dots mark the local minimum of the potential at which circular orbit is possible.
The star corresponds to the VISCO.
\label{fig: Veff APR rhoc2e14 v0-0.4}
}
\end{figure}
%%%%%%%%%%%%%%%%%%%%%%%%

In Fig.\ref{fig: Veff APR rhoc2e14 v0-0.4} we plot $V_{\rm eff}$ for a highly DM-enveloped
model with different value of $\ell_0$. Notice that a particle's radial motion is represented
by $E={\rm constant}$ line and $E>V_{\rm eff}$ corresponds to the allowed region of motion.
A bound orbit must have $E<0$. Minima of $V_{\rm eff}$ are radius of circular orbits.
Shaded-regions are the core (dark-colored. Mixture of baryon and DM) 
and the envelope (light-colored. Pure DM). The radius of the star $R/M=4.31$
is smaller than VISCO of $R/M=6$.  In vacuum region it is well-known
	that one stable circular orbit exists for $\ell_0>\sqrt{12}M$. There is another stable
	circular orbit in the DM-envelope as far as $\ell_0$ is not so large ($\ell_0\sim 4M$).
	Moreover we have a circular orbit in the envelope even with $\ell_0<\sqrt{12}M$.

For this particular model there are two separated region where stable circular orbit
is allowed. One is in the vacuum region which terminates at $r=6M$. The other is
in the DM-envelop. These regions are detached. Thus a thin accretion disk around
the star may have a gap at around the surface of the DM-envelope. The inner part of the disk
seems to extends down to the surface of the core.

To see where stable circular orbits are allowed we look at a criterion which results from
the curvature of $V_{\rm eff}$. If $d^2V_{\rm eff}/dr^2$ at an extremum of $V_{\rm eff}$ is
positive, the orbit is stable. The second derivative is proportional to the radial epicyclic frequency
squared.
	In our case the first and second derivatives of $V_{\rm eff}$ are,
	\begin{eqnarray}
		\frac{dV_{\rm eff}}{dr} &=& e^{2\nu}\left[\nu_{,r}\left(1+\frac{\ell_0^2}{r^2}\right)-\frac{\ell_0^2}{r^3}\right]\\
		\frac{d^2V_{\rm eff}}{dr^2} &=& 2e^{2\nu}\nu_{,r}\left[\nu_{,r}\left(1+\frac{\ell_0^2}{r^2}\right)-\frac{\ell_0^2}{r^3}\right]\nonumber\\
		&&+ e^{2\nu}\left[\nu_{,rr}\left(1+\frac{\ell_0^2}{r^2}\right)-\frac{2\ell_0^2\nu_{,r}}{r^3}
		+ \frac{3\ell_0^2}{r^4}\right],
	\end{eqnarray}
	where $\nu_{,r}$ and $\nu_{,rr}$ represent the first and second derivatives of $\nu$ with respect to $r$.
	The extrema of the first derivative are radii of circular orbits which satisfy,
	\begin{equation}
		\ell_0^2 = \frac{r^3\nu_{,r}}{1-r\nu_{,r}}.
		\label{eq: circular orbit}
	\end{equation}
	
	By nullifying the second derivative and using Eq.(\ref{eq: circular orbit}) 
	we obtain an equation for critical radius at which a circular orbit is an inflexion point of $V$,
	\begin{equation}
		F \equiv \nu_{,rr} - 2(\nu_{,r})^2 + \frac{3}{r}\nu_{,r} = 0
		\label{eq: critical radius}
	\end{equation}

%%%%%%%%%%%%%%%% Fig. F and E %%%%
\begin{figure}[hptb]
\includegraphics[scale=0.55]{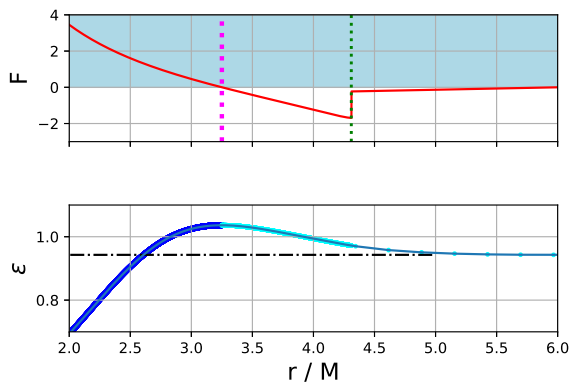}
\caption{Criterion $F$ (Eq.(\ref{eq: critical radius})) and specific energy $\epsilon$
(Eq.(\ref{eq: specific e and l})) as functions of radial coordinate. The equilibrium model
is the same as Fig.\ref{fig: Veff APR rhoc2e14 v0-0.4}.
[Upper panel] Criterion $F$ as a function of radial coordinate (solid). 
The shaded area corresponds to the allowed region
of a circular orbit. Thick dotted vertical line marks the inflexion point where $F$ vanishes.
The thin dotted vertical line marks the surface of the star which is the boundary between
DM envelope and the vacuum. 
[Lower panel] Specific energy $\epsilon$ of the circular orbits. Thick portion of the curve
corresponds to stable circular orbits. 
The horizontal dash-dotted
line is the energy at the vacuum last stable orbit ($r=6M$).
\label{fig: circular E and F rhoc2e14 v0-0.4}
}
\end{figure}
%%%%%%%%%%%%%%%%%%%%%%%
The upper panel in Fig.\ref{fig: circular E and F rhoc2e14 v0-0.4} shows $F$ of Eq.(\ref{eq: critical radius})
for a DM-enveloped configuration in Fig.\ref{fig: Veff APR rhoc2e14 v0-0.4}. 
Positive value of $F$ (shaded area) means a circular orbit 
is stable. Notice that the curve crosses zero at $r=6M$, which is the vacuum innermost stable circular
orbit (VISCO) around the mass $M$. The location of the surface of the DM envelope is marked
by the thin vertical dotted line, which is inside VISCO. The thick vertical dotted line marks the position
of the critical point inside the DM envelope where the radial stability of circular orbit changes.
On the left of the thick dotted line, another region of stable circular orbit exists.
As is seen in the lower panel of the figure, the specific energy the orbit 
in the range of $2.6<r/M<6$ exceeds that of VISCO. That means a thin circular accretion disk
may be truncated there, since the energy of the disk matter must decrease as it accretes inward.

%%%%%%%%%%%%%%%%% Veff %%%%
\begin{figure}
\includegraphics[scale=0.55]{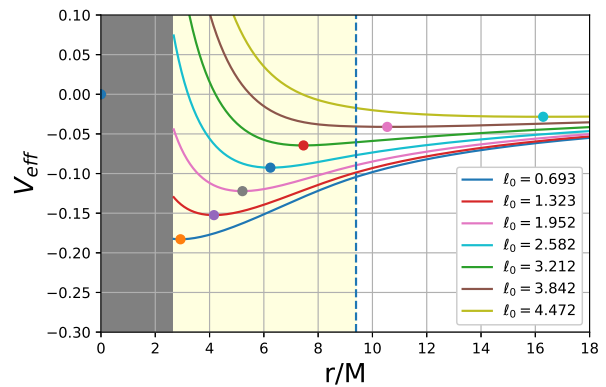}
\caption{Same as Fig.\ref{fig: Veff APR rhoc2e14 v0-0.4} except 
$v_0=0.1$.
The gravitational mass $M=2.12M_\odot$, baryonic mass $m_{_B}=1.37\times 10^{-2}M_\odot$,
DM rest mass $M_{_{DM}}=1.80M_\odot$. The radius of the core of baryon and DM mixture
is $R_{in}/M=2.68$ and the radius of the DM envelope $R_{DM}/M=9.40$.
\label{fig: Veff APR rhoc2e14 v0-0.1}
}
\end{figure}
\begin{figure}[hptb]
\includegraphics[scale=0.55]{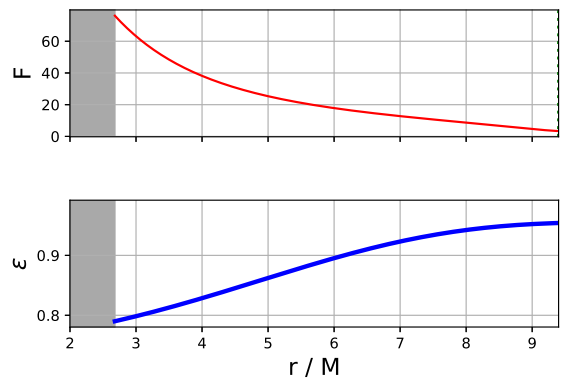}
\caption{Same as Fig.\ref{fig: circular E and F rhoc2e14 v0-0.4}
Except that $v_0=0.1$. The range of $r$ displayed is within
the star.
\label{fig: circular E and F rhoc2e14 v0-0.1}
}
\end{figure}

%%%%%%%%%%%%%%%%%%%%%%%%
In Fig.\ref{fig: Veff APR rhoc2e14 v0-0.1} effective potential is plotted for less 
DM-enveloped cases. We have 
$v_0=0.1$ with $\rho_c$ being fixed. No VISCO exists in this model.
We have only one circular orbit for $\ell_0/M\le 0.63$. A thin accretion disk is
expected to extend down to the core, where thin boundary layer may form.
Corresponding criterion $F$ and specific energy $\epsilon$ is found in 
Fig.\ref{fig: circular E and F rhoc2e14 v0-0.1}. We see that in the star $\epsilon$
is a monotonic function of $r$ and therefore an accretion disk may extends down
to the core without having a gap.
%%%%%%%%%%%%%%%% Fig. F and E %%%%
%%%%%%%%%%%%%%%%%%%%%%%

\begin{figure}[htpb]
\includegraphics[scale=0.55]{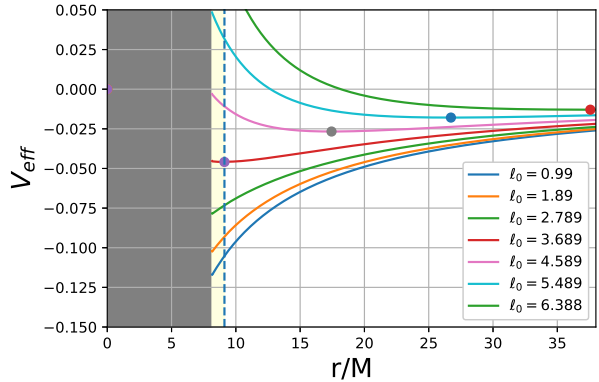}
\caption{Same as Fig.\ref{fig: Veff APR rhoc2e14 v0-0.1} except 
$\rho_c=7\times 10^{14}{\rm g}{\rm cm}^{-3}$.
The gravitational mass $M=0.84M_\odot$, baryonic mass $M_{_B}=0.682M_\odot$,
DM rest mass $M_{_{DM}}=0.214M_\odot$. The radius of the core of baryon and DM mixture
is $R_{in}/M=8.15$ and the radius of the DM envelope $R_{DM}/M=9.11$.
\label{fig: Veff APR rhoc7e14 v0-0.1}
}
\end{figure}

In Fig.\ref{fig: Veff APR rhoc7e14 v0-0.1} we also change $\rho_c$ as $7\times 10^{14}{\rm g}{\rm cm}^{-3}$.
The DM-envelope is further reduced. It is noted that baryonic mass exceeds that of DM for this case.
A single stable circular orbit exists for each $\ell_0$ whose radius extends down to
the surface of the core.

%%%%%%%%%%%%%%%%% Veff %%%%
%%%%%%%%%%%%%%%%%%%%%%%%

%%++++++++++++++++++++++++++++++++++++++++++++++
\subsection{Summary}
We explore characteristics of compact stars which is composed of
completely degenerate two species of matter, i.e., baryons and dark Fermions.
Interaction of two species are neglected. We solve TOV system of equations
for multi-component matter. The equilibrium models are parametrized by baryon
central density $\rho_c$ and Fermi momentum squared $v_0$ of dark Fermion 
at the center. Equilibrium is classified as baryon-enveloped, which has a core composed of 
mixture of baryons and dark Fermions and has an envelope composed solely of
baryons. When the envelope is entirely composed of DM, it is called DM-enveloped.
The former may be regarded as DANS and the latter
may be regarded as a generalized dark star. 
Since an equilibrium state is characterized by two parameters, we utilize Sorkin's
general criterion to investigate critical mass
of an equilibrium sequence beyond which the star becomes unstable.
We see that the baryon-enveloped and DM-enveloped classes have their own critical
masses determined mainly by the dominant component in their envelope.
For a typical parameter sets, we see that a balanced mixture of two species may
lead to smaller masses than if the either of the species is dominant.
This, however, depends on the mass and the strength of self-interaction of dark Fermion.

Another property of equilibrium investigated is the existence of stable circular orbits
around the star. Especially interesting is that the envelope may allow geodetic motion
of baryonic gases for DM-enveloped stars. We show that for a highly DM-enveloped
case circular orbits in the vacuum region terminates at VISCO, but another region
of stable circular orbits are possible in the DM envelope. For more mildly DM-enveloped
stars, we have a single circular orbit for each specific angular momentum down
to the baryon-mixed core, at which geodetic motion is not possible. In this case
a thin accretion disk that forms around the star does not have a gap and extends
down to the core. The existence or non-existence of the gap and the redshift
resulting from the gravity in the DM-envelope may affect the spectrum of disk
emission and may be an observable signature of these stars. Modeling of
these emissions may be an interesting astrophysical application of our model.

\begin{acknowledgments}
SY was supported by JSPS Grant-in-Aid for Scientific
Research(C) 18K03641\end{acknowledgments}

%\nocite{*}

\bibliography{dansref}

\providecommand{\noopsort}[1]{}\providecommand{\singleletter}[1]{#1}%
\begin{thebibliography}{42}%
\makeatletter
\providecommand \@ifxundefined [1]{%
 \@ifx{#1\undefined}
}%
\providecommand \@ifnum [1]{%
 \ifnum #1\expandafter \@firstoftwo
 \else \expandafter \@secondoftwo
 \fi
}%
\providecommand \@ifx [1]{%
 \ifx #1\expandafter \@firstoftwo
 \else \expandafter \@secondoftwo
 \fi
}%
\providecommand \natexlab [1]{#1}%
\providecommand \enquote  [1]{``#1''}%
\providecommand \bibnamefont  [1]{#1}%
\providecommand \bibfnamefont [1]{#1}%
\providecommand \citenamefont [1]{#1}%
\providecommand \href@noop [0]{\@secondoftwo}%
\providecommand \href [0]{\begingroup \@sanitize@url \@href}%
\providecommand \@href[1]{\@@startlink{#1}\@@href}%
\providecommand \@@href[1]{\endgroup#1\@@endlink}%
\providecommand \@sanitize@url [0]{\catcode `\\12\catcode `\$12\catcode
  `\&12\catcode `\#12\catcode `\^12\catcode `\_12\catcode `\%12\relax}%
\providecommand \@@startlink[1]{}%
\providecommand \@@endlink[0]{}%
\providecommand \url  [0]{\begingroup\@sanitize@url \@url }%
\providecommand \@url [1]{\endgroup\@href {#1}{\urlprefix }}%
\providecommand \urlprefix  [0]{URL }%
\providecommand \Eprint [0]{\href }%
\providecommand \doibase [0]{https://doi.org/}%
\providecommand \selectlanguage [0]{\@gobble}%
\providecommand \bibinfo  [0]{\@secondoftwo}%
\providecommand \bibfield  [0]{\@secondoftwo}%
\providecommand \translation [1]{[#1]}%
\providecommand \BibitemOpen [0]{}%
\providecommand \bibitemStop [0]{}%
\providecommand \bibitemNoStop [0]{.\EOS\space}%
\providecommand \EOS [0]{\spacefactor3000\relax}%
\providecommand \BibitemShut  [1]{\csname bibitem#1\endcsname}%
\let\auto@bib@innerbib\@empty
%</preamble>
\bibitem [{\citenamefont {{Zwicky}}(1933)}]{1933AcHPh...6..110Z}%
  \BibitemOpen
  \bibfield  {author} {\bibinfo {author} {\bibfnamefont {F.}~\bibnamefont
  {{Zwicky}}},\ }\bibfield  {title} {\bibinfo {title} {{Die Rotverschiebung von
  extragalaktischen Nebeln}},\ }\href@noop {} {\bibfield  {journal} {\bibinfo
  {journal} {Helvetica Physica Acta}\ }\textbf {\bibinfo {volume} {6}},\
  \bibinfo {pages} {110} (\bibinfo {year} {1933})}\BibitemShut {NoStop}%
\bibitem [{\citenamefont {{Zwicky}}(1937)}]{1937ApJ....86..217Z}%
  \BibitemOpen
  \bibfield  {author} {\bibinfo {author} {\bibfnamefont {F.}~\bibnamefont
  {{Zwicky}}},\ }\bibfield  {title} {\bibinfo {title} {{On the Masses of
  Nebulae and of Clusters of Nebulae}},\ }\href
  {https://doi.org/10.1086/143864} {\bibfield  {journal} {\bibinfo  {journal}
  {\apj}\ }\textbf {\bibinfo {volume} {86}},\ \bibinfo {pages} {217} (\bibinfo
  {year} {1937})}\BibitemShut {NoStop}%
\bibitem [{\citenamefont {{Rubin}}\ \emph {et~al.}(1980)\citenamefont
  {{Rubin}}, \citenamefont {{Ford}},\ and\ \citenamefont
  {{Thonnard}}}]{1980ApJ...238..471R}%
  \BibitemOpen
  \bibfield  {author} {\bibinfo {author} {\bibfnamefont {V.~C.}\ \bibnamefont
  {{Rubin}}}, \bibinfo {author} {\bibfnamefont {J.}~\bibnamefont {{Ford}},
  \bibfnamefont {W.~K.}},\ and\ \bibinfo {author} {\bibfnamefont
  {N.}~\bibnamefont {{Thonnard}}},\ }\bibfield  {title} {\bibinfo {title}
  {{Rotational properties of 21 SC galaxies with a large range of luminosities
  and radii, from NGC 4605 (R=4kpc) to UGC 2885 (R=122kpc).}},\ }\href
  {https://doi.org/10.1086/158003} {\bibfield  {journal} {\bibinfo  {journal}
  {\apj}\ }\textbf {\bibinfo {volume} {238}},\ \bibinfo {pages} {471} (\bibinfo
  {year} {1980})}\BibitemShut {NoStop}%
\bibitem [{\citenamefont {{Clowe}}\ \emph {et~al.}(2006)\citenamefont
  {{Clowe}}, \citenamefont {{Brada{\v{c}}}}, \citenamefont {{Gonzalez}},
  \citenamefont {{Markevitch}}, \citenamefont {{Randall}}, \citenamefont
  {{Jones}},\ and\ \citenamefont {{Zaritsky}}}]{2006ApJ...648L.109C}%
  \BibitemOpen
  \bibfield  {author} {\bibinfo {author} {\bibfnamefont {D.}~\bibnamefont
  {{Clowe}}}, \bibinfo {author} {\bibfnamefont {M.}~\bibnamefont
  {{Brada{\v{c}}}}}, \bibinfo {author} {\bibfnamefont {A.~H.}\ \bibnamefont
  {{Gonzalez}}}, \bibinfo {author} {\bibfnamefont {M.}~\bibnamefont
  {{Markevitch}}}, \bibinfo {author} {\bibfnamefont {S.~W.}\ \bibnamefont
  {{Randall}}}, \bibinfo {author} {\bibfnamefont {C.}~\bibnamefont {{Jones}}},\
  and\ \bibinfo {author} {\bibfnamefont {D.}~\bibnamefont {{Zaritsky}}},\
  }\bibfield  {title} {\bibinfo {title} {{A Direct Empirical Proof of the
  Existence of Dark Matter}},\ }\href {https://doi.org/10.1086/508162}
  {\bibfield  {journal} {\bibinfo  {journal} {\apjl}\ }\textbf {\bibinfo
  {volume} {648}},\ \bibinfo {pages} {L109} (\bibinfo {year} {2006})},\ \Eprint
  {https://arxiv.org/abs/astro-ph/0608407} {arXiv:astro-ph/0608407 [astro-ph]}
  \BibitemShut {NoStop}%
\bibitem [{\citenamefont {{Springel}}\ \emph {et~al.}(2006)\citenamefont
  {{Springel}}, \citenamefont {{Frenk}},\ and\ \citenamefont
  {{White}}}]{2006Natur.440.1137S}%
  \BibitemOpen
  \bibfield  {author} {\bibinfo {author} {\bibfnamefont {V.}~\bibnamefont
  {{Springel}}}, \bibinfo {author} {\bibfnamefont {C.~S.}\ \bibnamefont
  {{Frenk}}},\ and\ \bibinfo {author} {\bibfnamefont {S.~D.~M.}\ \bibnamefont
  {{White}}},\ }\bibfield  {title} {\bibinfo {title} {{The large-scale
  structure of the Universe}},\ }\href {https://doi.org/10.1038/nature04805}
  {\bibfield  {journal} {\bibinfo  {journal} {\nat}\ }\textbf {\bibinfo
  {volume} {440}},\ \bibinfo {pages} {1137} (\bibinfo {year} {2006})},\ \Eprint
  {https://arxiv.org/abs/astro-ph/0604561} {arXiv:astro-ph/0604561 [astro-ph]}
  \BibitemShut {NoStop}%
\bibitem [{\citenamefont {{Frenk}}\ and\ \citenamefont
  {{White}}(2012)}]{2012AnP...524..507F}%
  \BibitemOpen
  \bibfield  {author} {\bibinfo {author} {\bibfnamefont {C.~S.}\ \bibnamefont
  {{Frenk}}}\ and\ \bibinfo {author} {\bibfnamefont {S.~D.~M.}\ \bibnamefont
  {{White}}},\ }\bibfield  {title} {\bibinfo {title} {{Dark matter and cosmic
  structure}},\ }\href {https://doi.org/10.1002/andp.201200212} {\bibfield
  {journal} {\bibinfo  {journal} {Annalen der Physik}\ }\textbf {\bibinfo
  {volume} {524}},\ \bibinfo {pages} {507} (\bibinfo {year} {2012})},\ \Eprint
  {https://arxiv.org/abs/1210.0544} {arXiv:1210.0544 [astro-ph.CO]}
  \BibitemShut {NoStop}%
\bibitem [{\citenamefont {{Moore}}(1994)}]{1994Natur.370..629M}%
  \BibitemOpen
  \bibfield  {author} {\bibinfo {author} {\bibfnamefont {B.}~\bibnamefont
  {{Moore}}},\ }\bibfield  {title} {\bibinfo {title} {{Evidence against
  dissipation-less dark matter from observations of galaxy haloes}},\ }\href
  {https://doi.org/10.1038/370629a0} {\bibfield  {journal} {\bibinfo  {journal}
  {\nat}\ }\textbf {\bibinfo {volume} {370}},\ \bibinfo {pages} {629} (\bibinfo
  {year} {1994})}\BibitemShut {NoStop}%
\bibitem [{\citenamefont {{Flores}}\ and\ \citenamefont
  {{Primack}}(1994)}]{1994ApJ...427L...1F}%
  \BibitemOpen
  \bibfield  {author} {\bibinfo {author} {\bibfnamefont {R.~A.}\ \bibnamefont
  {{Flores}}}\ and\ \bibinfo {author} {\bibfnamefont {J.~R.}\ \bibnamefont
  {{Primack}}},\ }\bibfield  {title} {\bibinfo {title} {{Observational and
  Theoretical Constraints on Singular Dark Matter Halos}},\ }\href
  {https://doi.org/10.1086/187350} {\bibfield  {journal} {\bibinfo  {journal}
  {\apjl}\ }\textbf {\bibinfo {volume} {427}},\ \bibinfo {pages} {L1} (\bibinfo
  {year} {1994})}\BibitemShut {NoStop}%
\bibitem [{\citenamefont {{Kauffmann}}\ \emph {et~al.}(1993)\citenamefont
  {{Kauffmann}}, \citenamefont {{White}},\ and\ \citenamefont
  {{Guiderdoni}}}]{1993MNRAS.264..201K}%
  \BibitemOpen
  \bibfield  {author} {\bibinfo {author} {\bibfnamefont {G.}~\bibnamefont
  {{Kauffmann}}}, \bibinfo {author} {\bibfnamefont {S.~D.~M.}\ \bibnamefont
  {{White}}},\ and\ \bibinfo {author} {\bibfnamefont {B.}~\bibnamefont
  {{Guiderdoni}}},\ }\bibfield  {title} {\bibinfo {title} {{The formation and
  evolution of galaxies within merging dark matter haloes.}},\ }\href
  {https://doi.org/10.1093/mnras/264.1.201} {\bibfield  {journal} {\bibinfo
  {journal} {\mnras}\ }\textbf {\bibinfo {volume} {264}},\ \bibinfo {pages}
  {201} (\bibinfo {year} {1993})}\BibitemShut {NoStop}%
\bibitem [{\citenamefont {{Klypin}}\ \emph {et~al.}(1999)\citenamefont
  {{Klypin}}, \citenamefont {{Kravtsov}}, \citenamefont {{Valenzuela}},\ and\
  \citenamefont {{Prada}}}]{1999ApJ...522...82K}%
  \BibitemOpen
  \bibfield  {author} {\bibinfo {author} {\bibfnamefont {A.}~\bibnamefont
  {{Klypin}}}, \bibinfo {author} {\bibfnamefont {A.~V.}\ \bibnamefont
  {{Kravtsov}}}, \bibinfo {author} {\bibfnamefont {O.}~\bibnamefont
  {{Valenzuela}}},\ and\ \bibinfo {author} {\bibfnamefont {F.}~\bibnamefont
  {{Prada}}},\ }\bibfield  {title} {\bibinfo {title} {{Where Are the Missing
  Galactic Satellites?}},\ }\href {https://doi.org/10.1086/307643} {\bibfield
  {journal} {\bibinfo  {journal} {\apj}\ }\textbf {\bibinfo {volume} {522}},\
  \bibinfo {pages} {82} (\bibinfo {year} {1999})},\ \Eprint
  {https://arxiv.org/abs/astro-ph/9901240} {arXiv:astro-ph/9901240 [astro-ph]}
  \BibitemShut {NoStop}%
\bibitem [{\citenamefont {{Moore}}\ \emph {et~al.}(1999)\citenamefont
  {{Moore}}, \citenamefont {{Ghigna}}, \citenamefont {{Governato}},
  \citenamefont {{Lake}}, \citenamefont {{Quinn}}, \citenamefont {{Stadel}},\
  and\ \citenamefont {{Tozzi}}}]{1999ApJ...524L..19M}%
  \BibitemOpen
  \bibfield  {author} {\bibinfo {author} {\bibfnamefont {B.}~\bibnamefont
  {{Moore}}}, \bibinfo {author} {\bibfnamefont {S.}~\bibnamefont {{Ghigna}}},
  \bibinfo {author} {\bibfnamefont {F.}~\bibnamefont {{Governato}}}, \bibinfo
  {author} {\bibfnamefont {G.}~\bibnamefont {{Lake}}}, \bibinfo {author}
  {\bibfnamefont {T.}~\bibnamefont {{Quinn}}}, \bibinfo {author} {\bibfnamefont
  {J.}~\bibnamefont {{Stadel}}},\ and\ \bibinfo {author} {\bibfnamefont
  {P.}~\bibnamefont {{Tozzi}}},\ }\bibfield  {title} {\bibinfo {title} {{Dark
  Matter Substructure within Galactic Halos}},\ }\href
  {https://doi.org/10.1086/312287} {\bibfield  {journal} {\bibinfo  {journal}
  {\apjl}\ }\textbf {\bibinfo {volume} {524}},\ \bibinfo {pages} {L19}
  (\bibinfo {year} {1999})},\ \Eprint {https://arxiv.org/abs/astro-ph/9907411}
  {arXiv:astro-ph/9907411 [astro-ph]} \BibitemShut {NoStop}%
\bibitem [{\citenamefont {{Boylan-Kolchin}}\ \emph {et~al.}(2011)\citenamefont
  {{Boylan-Kolchin}}, \citenamefont {{Bullock}},\ and\ \citenamefont
  {{Kaplinghat}}}]{2011MNRAS.415L..40B}%
  \BibitemOpen
  \bibfield  {author} {\bibinfo {author} {\bibfnamefont {M.}~\bibnamefont
  {{Boylan-Kolchin}}}, \bibinfo {author} {\bibfnamefont {J.~S.}\ \bibnamefont
  {{Bullock}}},\ and\ \bibinfo {author} {\bibfnamefont {M.}~\bibnamefont
  {{Kaplinghat}}},\ }\bibfield  {title} {\bibinfo {title} {{Too big to fail?
  The puzzling darkness of massive Milky Way subhaloes}},\ }\href
  {https://doi.org/10.1111/j.1745-3933.2011.01074.x} {\bibfield  {journal}
  {\bibinfo  {journal} {\mnras}\ }\textbf {\bibinfo {volume} {415}},\ \bibinfo
  {pages} {L40} (\bibinfo {year} {2011})},\ \Eprint
  {https://arxiv.org/abs/1103.0007} {arXiv:1103.0007 [astro-ph.CO]}
  \BibitemShut {NoStop}%
\bibitem [{\citenamefont {{Vogelsberger}}\ \emph {et~al.}(2012)\citenamefont
  {{Vogelsberger}}, \citenamefont {{Zavala}},\ and\ \citenamefont
  {{Loeb}}}]{2012MNRAS.423.3740V}%
  \BibitemOpen
  \bibfield  {author} {\bibinfo {author} {\bibfnamefont {M.}~\bibnamefont
  {{Vogelsberger}}}, \bibinfo {author} {\bibfnamefont {J.}~\bibnamefont
  {{Zavala}}},\ and\ \bibinfo {author} {\bibfnamefont {A.}~\bibnamefont
  {{Loeb}}},\ }\bibfield  {title} {\bibinfo {title} {{Subhaloes in
  self-interacting galactic dark matter haloes}},\ }\href
  {https://doi.org/10.1111/j.1365-2966.2012.21182.x} {\bibfield  {journal}
  {\bibinfo  {journal} {\mnras}\ }\textbf {\bibinfo {volume} {423}},\ \bibinfo
  {pages} {3740} (\bibinfo {year} {2012})},\ \Eprint
  {https://arxiv.org/abs/1201.5892} {arXiv:1201.5892 [astro-ph.CO]}
  \BibitemShut {NoStop}%
\bibitem [{\citenamefont {{Rocha}}\ \emph {et~al.}(2013)\citenamefont
  {{Rocha}}, \citenamefont {{Peter}}, \citenamefont {{Bullock}}, \citenamefont
  {{Kaplinghat}}, \citenamefont {{Garrison-Kimmel}}, \citenamefont
  {{O{\~n}orbe}},\ and\ \citenamefont {{Moustakas}}}]{2013MNRAS.430...81R}%
  \BibitemOpen
  \bibfield  {author} {\bibinfo {author} {\bibfnamefont {M.}~\bibnamefont
  {{Rocha}}}, \bibinfo {author} {\bibfnamefont {A.~H.~G.}\ \bibnamefont
  {{Peter}}}, \bibinfo {author} {\bibfnamefont {J.~S.}\ \bibnamefont
  {{Bullock}}}, \bibinfo {author} {\bibfnamefont {M.}~\bibnamefont
  {{Kaplinghat}}}, \bibinfo {author} {\bibfnamefont {S.}~\bibnamefont
  {{Garrison-Kimmel}}}, \bibinfo {author} {\bibfnamefont {J.}~\bibnamefont
  {{O{\~n}orbe}}},\ and\ \bibinfo {author} {\bibfnamefont {L.~A.}\ \bibnamefont
  {{Moustakas}}},\ }\bibfield  {title} {\bibinfo {title} {{Cosmological
  simulations with self-interacting dark matter - I. Constant-density cores and
  substructure}},\ }\href {https://doi.org/10.1093/mnras/sts514} {\bibfield
  {journal} {\bibinfo  {journal} {\mnras}\ }\textbf {\bibinfo {volume} {430}},\
  \bibinfo {pages} {81} (\bibinfo {year} {2013})},\ \Eprint
  {https://arxiv.org/abs/1208.3025} {arXiv:1208.3025 [astro-ph.CO]}
  \BibitemShut {NoStop}%
\bibitem [{\citenamefont {{Zavala}}\ \emph {et~al.}(2013)\citenamefont
  {{Zavala}}, \citenamefont {{Vogelsberger}},\ and\ \citenamefont
  {{Walker}}}]{2013MNRAS.431L..20Z}%
  \BibitemOpen
  \bibfield  {author} {\bibinfo {author} {\bibfnamefont {J.}~\bibnamefont
  {{Zavala}}}, \bibinfo {author} {\bibfnamefont {M.}~\bibnamefont
  {{Vogelsberger}}},\ and\ \bibinfo {author} {\bibfnamefont {M.~G.}\
  \bibnamefont {{Walker}}},\ }\bibfield  {title} {\bibinfo {title}
  {{Constraining self-interacting dark matter with the Milky way's dwarf
  spheroidals.}},\ }\href {https://doi.org/10.1093/mnrasl/sls053} {\bibfield
  {journal} {\bibinfo  {journal} {\mnras}\ }\textbf {\bibinfo {volume} {431}},\
  \bibinfo {pages} {L20} (\bibinfo {year} {2013})},\ \Eprint
  {https://arxiv.org/abs/1211.6426} {arXiv:1211.6426 [astro-ph.CO]}
  \BibitemShut {NoStop}%
\bibitem [{\citenamefont {{Peter}}\ \emph {et~al.}(2013)\citenamefont
  {{Peter}}, \citenamefont {{Rocha}}, \citenamefont {{Bullock}},\ and\
  \citenamefont {{Kaplinghat}}}]{2013MNRAS.430..105P}%
  \BibitemOpen
  \bibfield  {author} {\bibinfo {author} {\bibfnamefont {A.~H.~G.}\
  \bibnamefont {{Peter}}}, \bibinfo {author} {\bibfnamefont {M.}~\bibnamefont
  {{Rocha}}}, \bibinfo {author} {\bibfnamefont {J.~S.}\ \bibnamefont
  {{Bullock}}},\ and\ \bibinfo {author} {\bibfnamefont {M.}~\bibnamefont
  {{Kaplinghat}}},\ }\bibfield  {title} {\bibinfo {title} {{Cosmological
  simulations with self-interacting dark matter - II. Halo shapes versus
  observations}},\ }\href {https://doi.org/10.1093/mnras/sts535} {\bibfield
  {journal} {\bibinfo  {journal} {\mnras}\ }\textbf {\bibinfo {volume} {430}},\
  \bibinfo {pages} {105} (\bibinfo {year} {2013})},\ \Eprint
  {https://arxiv.org/abs/1208.3026} {arXiv:1208.3026 [astro-ph.CO]}
  \BibitemShut {NoStop}%
\bibitem [{\citenamefont {{Zurek}}(2014)}]{2014PhR...537...91Z}%
  \BibitemOpen
  \bibfield  {author} {\bibinfo {author} {\bibfnamefont {K.~M.}\ \bibnamefont
  {{Zurek}}},\ }\bibfield  {title} {\bibinfo {title} {{Asymmetric Dark Matter:
  Theories, signatures, and constraints}},\ }\href
  {https://doi.org/10.1016/j.physrep.2013.12.001} {\bibfield  {journal}
  {\bibinfo  {journal} {\physrep}\ }\textbf {\bibinfo {volume} {537}},\
  \bibinfo {pages} {91} (\bibinfo {year} {2014})},\ \Eprint
  {https://arxiv.org/abs/1308.0338} {arXiv:1308.0338 [hep-ph]} \BibitemShut
  {NoStop}%
\bibitem [{\citenamefont {{Yuan}}\ \emph {et~al.}(2004)\citenamefont {{Yuan}},
  \citenamefont {{Narayan}},\ and\ \citenamefont
  {{Rees}}}]{2004ApJ...606.1112Y}%
  \BibitemOpen
  \bibfield  {author} {\bibinfo {author} {\bibfnamefont {Y.-F.}\ \bibnamefont
  {{Yuan}}}, \bibinfo {author} {\bibfnamefont {R.}~\bibnamefont {{Narayan}}},\
  and\ \bibinfo {author} {\bibfnamefont {M.~J.}\ \bibnamefont {{Rees}}},\
  }\bibfield  {title} {\bibinfo {title} {{Constraining Alternate Models of
  Black Holes: Type I X-Ray Bursts on Accreting Fermion-Fermion and
  Boson-Fermion Stars}},\ }\href {https://doi.org/10.1086/383185} {\bibfield
  {journal} {\bibinfo  {journal} {\apj}\ }\textbf {\bibinfo {volume} {606}},\
  \bibinfo {pages} {1112} (\bibinfo {year} {2004})},\ \Eprint
  {https://arxiv.org/abs/astro-ph/0401549} {arXiv:astro-ph/0401549 [astro-ph]}
  \BibitemShut {NoStop}%
\bibitem [{Note1()}]{Note1}%
  \BibitemOpen
  \bibinfo {note} {It should be noted that if DM particles are not asymmetric
  as in popular WIMPs (weakly interacting massive particles) or axions, they
  may annihilate each other and heat a neutron star after they are captured by
  it. These possibilities have been tested by studying the cooling history of
  neutron stars (see, e.g., \cite {PhysRevD.77.023006}).}\BibitemShut {Stop}%
\bibitem [{\citenamefont {{Kouvaris}}\ and\ \citenamefont
  {{Nielsen}}(2015)}]{2015PhRvD..92f3526K}%
  \BibitemOpen
  \bibfield  {author} {\bibinfo {author} {\bibfnamefont {C.}~\bibnamefont
  {{Kouvaris}}}\ and\ \bibinfo {author} {\bibfnamefont {N.~G.}\ \bibnamefont
  {{Nielsen}}},\ }\bibfield  {title} {\bibinfo {title} {{Asymmetric dark matter
  stars}},\ }\href {https://doi.org/10.1103/PhysRevD.92.063526} {\bibfield
  {journal} {\bibinfo  {journal} {\prd}\ }\textbf {\bibinfo {volume} {92}},\
  \bibinfo {eid} {063526} (\bibinfo {year} {2015})},\ \Eprint
  {https://arxiv.org/abs/1507.00959} {arXiv:1507.00959 [hep-ph]} \BibitemShut
  {NoStop}%
\bibitem [{\citenamefont {{Maselli}}\ \emph
  {et~al.}(2017{\natexlab{a}})\citenamefont {{Maselli}}, \citenamefont
  {{Pnigouras}}, \citenamefont {{Nielsen}}, \citenamefont {{Kouvaris}},\ and\
  \citenamefont {{Kokkotas}}}]{2017PhRvD..96b3005M}%
  \BibitemOpen
  \bibfield  {author} {\bibinfo {author} {\bibfnamefont {A.}~\bibnamefont
  {{Maselli}}}, \bibinfo {author} {\bibfnamefont {P.}~\bibnamefont
  {{Pnigouras}}}, \bibinfo {author} {\bibfnamefont {N.~G.}\ \bibnamefont
  {{Nielsen}}}, \bibinfo {author} {\bibfnamefont {C.}~\bibnamefont
  {{Kouvaris}}},\ and\ \bibinfo {author} {\bibfnamefont {K.~D.}\ \bibnamefont
  {{Kokkotas}}},\ }\bibfield  {title} {\bibinfo {title} {{Dark stars:
  Gravitational and electromagnetic observables}},\ }\href
  {https://doi.org/10.1103/PhysRevD.96.023005} {\bibfield  {journal} {\bibinfo
  {journal} {\prd}\ }\textbf {\bibinfo {volume} {96}},\ \bibinfo {eid} {023005}
  (\bibinfo {year} {2017}{\natexlab{a}})},\ \Eprint
  {https://arxiv.org/abs/1704.07286} {arXiv:1704.07286 [astro-ph.HE]}
  \BibitemShut {NoStop}%
\bibitem [{\citenamefont {Leung}\ \emph {et~al.}(2011)\citenamefont {Leung},
  \citenamefont {Chu},\ and\ \citenamefont {Lin}}]{PhysRevD.84.107301}%
  \BibitemOpen
  \bibfield  {author} {\bibinfo {author} {\bibfnamefont {S.-C.}\ \bibnamefont
  {Leung}}, \bibinfo {author} {\bibfnamefont {M.-C.}\ \bibnamefont {Chu}},\
  and\ \bibinfo {author} {\bibfnamefont {L.-M.}\ \bibnamefont {Lin}},\
  }\bibfield  {title} {\bibinfo {title} {Dark-matter admixed neutron stars},\
  }\href {https://doi.org/10.1103/PhysRevD.84.107301} {\bibfield  {journal}
  {\bibinfo  {journal} {Phys. Rev. D}\ }\textbf {\bibinfo {volume} {84}},\
  \bibinfo {pages} {107301} (\bibinfo {year} {2011})}\BibitemShut {NoStop}%
\bibitem [{\citenamefont {{Leung}}\ \emph {et~al.}(2019)\citenamefont
  {{Leung}}, \citenamefont {{Zha}}, \citenamefont {{Chu}}, \citenamefont
  {{Lin}},\ and\ \citenamefont {{Nomoto}}}]{2019ApJ...884....9L}%
  \BibitemOpen
  \bibfield  {author} {\bibinfo {author} {\bibfnamefont {S.-C.}\ \bibnamefont
  {{Leung}}}, \bibinfo {author} {\bibfnamefont {S.}~\bibnamefont {{Zha}}},
  \bibinfo {author} {\bibfnamefont {M.-C.}\ \bibnamefont {{Chu}}}, \bibinfo
  {author} {\bibfnamefont {L.-M.}\ \bibnamefont {{Lin}}},\ and\ \bibinfo
  {author} {\bibfnamefont {K.}~\bibnamefont {{Nomoto}}},\ }\bibfield  {title}
  {\bibinfo {title} {{Accretion-induced Collapse of Dark Matter Admixed White
  Dwarfs. I. Formation of Low-mass Neutron Stars}},\ }\href
  {https://doi.org/10.3847/1538-4357/ab3b5e} {\bibfield  {journal} {\bibinfo
  {journal} {\apj}\ }\textbf {\bibinfo {volume} {884}},\ \bibinfo {eid} {9}
  (\bibinfo {year} {2019})},\ \Eprint {https://arxiv.org/abs/1908.05102}
  {arXiv:1908.05102 [astro-ph.HE]} \BibitemShut {NoStop}%
\bibitem [{\citenamefont {Leung}\ \emph {et~al.}(2012)\citenamefont {Leung},
  \citenamefont {Chu},\ and\ \citenamefont {Lin}}]{PhysRevD.85.103528}%
  \BibitemOpen
  \bibfield  {author} {\bibinfo {author} {\bibfnamefont {S.-C.}\ \bibnamefont
  {Leung}}, \bibinfo {author} {\bibfnamefont {M.-C.}\ \bibnamefont {Chu}},\
  and\ \bibinfo {author} {\bibfnamefont {L.-M.}\ \bibnamefont {Lin}},\
  }\bibfield  {title} {\bibinfo {title} {Equilibrium structure and radial
  oscillations of dark matter admixed neutron stars},\ }\href
  {https://doi.org/10.1103/PhysRevD.85.103528} {\bibfield  {journal} {\bibinfo
  {journal} {Phys. Rev. D}\ }\textbf {\bibinfo {volume} {85}},\ \bibinfo
  {pages} {103528} (\bibinfo {year} {2012})}\BibitemShut {NoStop}%
\bibitem [{Note2()}]{Note2}%
  \BibitemOpen
  \bibinfo {note} {These objects are called 'double degenerate stars' in \cite
  {2008ChPhL..25.2378L}. We have slightly modified the term because it may be
  confused with binary stars whose components are degenerate
  stars.}\BibitemShut {Stop}%
\bibitem [{\citenamefont {{Ciarcelluti}}\ and\ \citenamefont
  {{Sandin}}(2011)}]{Ciarcelluti_Sandin2011}%
  \BibitemOpen
  \bibfield  {author} {\bibinfo {author} {\bibfnamefont {P.}~\bibnamefont
  {{Ciarcelluti}}}\ and\ \bibinfo {author} {\bibfnamefont {F.}~\bibnamefont
  {{Sandin}}},\ }\bibfield  {title} {\bibinfo {title} {{Have neutron stars a
  dark matter core?}},\ }\href {https://doi.org/10.1016/j.physletb.2010.11.021}
  {\bibfield  {journal} {\bibinfo  {journal} {Physics Letters B}\ }\textbf
  {\bibinfo {volume} {695}},\ \bibinfo {pages} {19} (\bibinfo {year} {2011})},\
  \Eprint {https://arxiv.org/abs/1005.0857} {arXiv:1005.0857 [astro-ph.HE]}
  \BibitemShut {NoStop}%
\bibitem [{\citenamefont {{Gresham}}\ and\ \citenamefont
  {{Zurek}}(2019)}]{Gresham_Zurek2019}%
  \BibitemOpen
  \bibfield  {author} {\bibinfo {author} {\bibfnamefont {M.~I.}\ \bibnamefont
  {{Gresham}}}\ and\ \bibinfo {author} {\bibfnamefont {K.~M.}\ \bibnamefont
  {{Zurek}}},\ }\bibfield  {title} {\bibinfo {title} {{Asymmetric dark stars
  and neutron star stability}},\ }\href
  {https://doi.org/10.1103/PhysRevD.99.083008} {\bibfield  {journal} {\bibinfo
  {journal} {\prd}\ }\textbf {\bibinfo {volume} {99}},\ \bibinfo {eid} {083008}
  (\bibinfo {year} {2019})},\ \Eprint {https://arxiv.org/abs/1809.08254}
  {arXiv:1809.08254 [astro-ph.CO]} \BibitemShut {NoStop}%
\bibitem [{\citenamefont {{Kodama}}\ and\ \citenamefont
  {{Yamada}}(1972)}]{Kodama_Yamada1972}%
  \BibitemOpen
  \bibfield  {author} {\bibinfo {author} {\bibfnamefont {T.}~\bibnamefont
  {{Kodama}}}\ and\ \bibinfo {author} {\bibfnamefont {M.}~\bibnamefont
  {{Yamada}}},\ }\bibfield  {title} {\bibinfo {title} {{Theory of Superdense
  Stars}},\ }\href {https://doi.org/10.1143/PTP.47.444} {\bibfield  {journal}
  {\bibinfo  {journal} {Progress of Theoretical Physics}\ }\textbf {\bibinfo
  {volume} {47}},\ \bibinfo {pages} {444} (\bibinfo {year} {1972})}\BibitemShut
  {NoStop}%
\bibitem [{\citenamefont {{Maselli}}\ \emph
  {et~al.}(2017{\natexlab{b}})\citenamefont {{Maselli}}, \citenamefont
  {{Pnigouras}}, \citenamefont {{Nielsen}}, \citenamefont {{Kouvaris}},\ and\
  \citenamefont {{Kokkotas}}}]{Maselli2017}%
  \BibitemOpen
  \bibfield  {author} {\bibinfo {author} {\bibfnamefont {A.}~\bibnamefont
  {{Maselli}}}, \bibinfo {author} {\bibfnamefont {P.}~\bibnamefont
  {{Pnigouras}}}, \bibinfo {author} {\bibfnamefont {N.~G.}\ \bibnamefont
  {{Nielsen}}}, \bibinfo {author} {\bibfnamefont {C.}~\bibnamefont
  {{Kouvaris}}},\ and\ \bibinfo {author} {\bibfnamefont {K.~D.}\ \bibnamefont
  {{Kokkotas}}},\ }\bibfield  {title} {\bibinfo {title} {{Dark stars:
  Gravitational and electromagnetic observables}},\ }\href
  {https://doi.org/10.1103/PhysRevD.96.023005} {\bibfield  {journal} {\bibinfo
  {journal} {\prd}\ }\textbf {\bibinfo {volume} {96}},\ \bibinfo {eid} {023005}
  (\bibinfo {year} {2017}{\natexlab{b}})},\ \Eprint
  {https://arxiv.org/abs/1704.07286} {arXiv:1704.07286 [astro-ph.HE]}
  \BibitemShut {NoStop}%
\bibitem [{\citenamefont {{Kouvaris}}(2012)}]{Kouvaris2012}%
  \BibitemOpen
  \bibfield  {author} {\bibinfo {author} {\bibfnamefont {C.}~\bibnamefont
  {{Kouvaris}}},\ }\bibfield  {title} {\bibinfo {title} {{Limits on
  Self-Interacting Dark Matter from Neutron Stars}},\ }\href
  {https://doi.org/10.1103/PhysRevLett.108.191301} {\bibfield  {journal}
  {\bibinfo  {journal} {\prl}\ }\textbf {\bibinfo {volume} {108}},\ \bibinfo
  {eid} {191301} (\bibinfo {year} {2012})},\ \Eprint
  {https://arxiv.org/abs/1111.4364} {arXiv:1111.4364 [astro-ph.CO]}
  \BibitemShut {NoStop}%
\bibitem [{Note3()}]{Note3}%
  \BibitemOpen
  \bibinfo {note} {Attractive self-interaction does not necessarily mean a star
  is unstable to gravitational collapse as shown in \cite {Gresham_Zurek2019}.
  We focus on the repulsive case for simplicity.}\BibitemShut {Stop}%
\bibitem [{\citenamefont {{Tulin}}\ and\ \citenamefont
  {{Yu}}(2018)}]{2018PhR...730....1T}%
  \BibitemOpen
  \bibfield  {author} {\bibinfo {author} {\bibfnamefont {S.}~\bibnamefont
  {{Tulin}}}\ and\ \bibinfo {author} {\bibfnamefont {H.-B.}\ \bibnamefont
  {{Yu}}},\ }\bibfield  {title} {\bibinfo {title} {{Dark matter
  self-interactions and small scale structure}},\ }\href
  {https://doi.org/10.1016/j.physrep.2017.11.004} {\bibfield  {journal}
  {\bibinfo  {journal} {Phys. Rep.}\ }\textbf {\bibinfo {volume} {730}},\
  \bibinfo {pages} {1} (\bibinfo {year} {2018})},\ \Eprint
  {https://arxiv.org/abs/1705.02358} {arXiv:1705.02358 [hep-ph]} \BibitemShut
  {NoStop}%
\bibitem [{\citenamefont {{Haensel}}\ and\ \citenamefont
  {{Potekhin}}(2004)}]{Haensel_Potekhin2004}%
  \BibitemOpen
  \bibfield  {author} {\bibinfo {author} {\bibfnamefont {P.}~\bibnamefont
  {{Haensel}}}\ and\ \bibinfo {author} {\bibfnamefont {A.~Y.}\ \bibnamefont
  {{Potekhin}}},\ }\bibfield  {title} {\bibinfo {title} {{Analytical
  representations of unified equations of state of neutron-star matter}},\
  }\href {https://doi.org/10.1051/0004-6361:20041722} {\bibfield  {journal}
  {\bibinfo  {journal} {Astron. Astrophys.}\ }\textbf {\bibinfo {volume}
  {428}},\ \bibinfo {pages} {191} (\bibinfo {year} {2004})},\ \Eprint
  {https://arxiv.org/abs/astro-ph/0408324} {arXiv:astro-ph/0408324 [astro-ph]}
  \BibitemShut {NoStop}%
\bibitem [{\citenamefont {{Potekhin}}\ and\ \citenamefont
  {{Chabrier}}(2018)}]{Potekhin_Chabrier2018}%
  \BibitemOpen
  \bibfield  {author} {\bibinfo {author} {\bibfnamefont {A.~Y.}\ \bibnamefont
  {{Potekhin}}}\ and\ \bibinfo {author} {\bibfnamefont {G.}~\bibnamefont
  {{Chabrier}}},\ }\bibfield  {title} {\bibinfo {title} {{Magnetic neutron star
  cooling and microphysics}},\ }\href
  {https://doi.org/10.1051/0004-6361/201731866} {\bibfield  {journal} {\bibinfo
   {journal} {Astron. Astrophys}\ }\textbf {\bibinfo {volume} {609}},\ \bibinfo
  {eid} {A74} (\bibinfo {year} {2018})},\ \Eprint
  {https://arxiv.org/abs/1711.07662} {arXiv:1711.07662 [astro-ph.HE]}
  \BibitemShut {NoStop}%
\bibitem [{\citenamefont {{Pandharipande}}\ and\ \citenamefont
  {{Ravenhall}}(1989)}]{FPS1989}%
  \BibitemOpen
  \bibfield  {author} {\bibinfo {author} {\bibfnamefont {V.~R.}\ \bibnamefont
  {{Pandharipande}}}\ and\ \bibinfo {author} {\bibfnamefont {D.~G.}\
  \bibnamefont {{Ravenhall}}},\ }\bibfield  {title} {\bibinfo {title} {{Hot
  Nuclear Matter}},\ }in\ \href@noop {} {\emph {\bibinfo {booktitle} {NATO
  Advanced Science Institutes (ASI) Series B}}},\ Vol.\ \bibinfo {volume}
  {205}\ (\bibinfo {year} {1989})\ p.\ \bibinfo {pages} {103}\BibitemShut
  {NoStop}%
\bibitem [{\citenamefont {{Douchin}}\ and\ \citenamefont
  {{Haensel}}(2001)}]{SLy2001}%
  \BibitemOpen
  \bibfield  {author} {\bibinfo {author} {\bibfnamefont {F.}~\bibnamefont
  {{Douchin}}}\ and\ \bibinfo {author} {\bibfnamefont {P.}~\bibnamefont
  {{Haensel}}},\ }\bibfield  {title} {\bibinfo {title} {{A unified equation of
  state of dense matter and neutron star structure}},\ }\href
  {https://doi.org/10.1051/0004-6361:20011402} {\bibfield  {journal} {\bibinfo
  {journal} {Astron. Astrophys.}\ }\textbf {\bibinfo {volume} {380}},\ \bibinfo
  {pages} {151} (\bibinfo {year} {2001})},\ \Eprint
  {https://arxiv.org/abs/astro-ph/0111092} {arXiv:astro-ph/0111092 [astro-ph]}
  \BibitemShut {NoStop}%
\bibitem [{\citenamefont {{Akmal}}\ \emph {et~al.}(1998)\citenamefont
  {{Akmal}}, \citenamefont {{Pandharipande}},\ and\ \citenamefont
  {{Ravenhall}}}]{APR1998}%
  \BibitemOpen
  \bibfield  {author} {\bibinfo {author} {\bibfnamefont {A.}~\bibnamefont
  {{Akmal}}}, \bibinfo {author} {\bibfnamefont {V.~R.}\ \bibnamefont
  {{Pandharipande}}},\ and\ \bibinfo {author} {\bibfnamefont {D.~G.}\
  \bibnamefont {{Ravenhall}}},\ }\bibfield  {title} {\bibinfo {title}
  {{Equation of state of nucleon matter and neutron star structure}},\ }\href
  {https://doi.org/10.1103/PhysRevC.58.1804} {\bibfield  {journal} {\bibinfo
  {journal} {\prc}\ }\textbf {\bibinfo {volume} {58}},\ \bibinfo {pages} {1804}
  (\bibinfo {year} {1998})},\ \Eprint {https://arxiv.org/abs/nucl-th/9804027}
  {arXiv:nucl-th/9804027 [nucl-th]} \BibitemShut {NoStop}%
\bibitem [{\citenamefont {{Sorkin}}(1982)}]{1982ApJ...257..847S}%
  \BibitemOpen
  \bibfield  {author} {\bibinfo {author} {\bibfnamefont {R.~D.}\ \bibnamefont
  {{Sorkin}}},\ }\bibfield  {title} {\bibinfo {title} {{A Stability Criterion
  for Many Parameter Equilibrium Families}},\ }\href
  {https://doi.org/10.1086/160034} {\bibfield  {journal} {\bibinfo  {journal}
  {\apj}\ }\textbf {\bibinfo {volume} {257}},\ \bibinfo {pages} {847} (\bibinfo
  {year} {1982})}\BibitemShut {NoStop}%
\bibitem [{\citenamefont {{Friedman}}\ \emph {et~al.}(1988)\citenamefont
  {{Friedman}}, \citenamefont {{Ipser}},\ and\ \citenamefont
  {{Sorkin}}}]{1988ApJ...325..722F}%
  \BibitemOpen
  \bibfield  {author} {\bibinfo {author} {\bibfnamefont {J.~L.}\ \bibnamefont
  {{Friedman}}}, \bibinfo {author} {\bibfnamefont {J.~R.}\ \bibnamefont
  {{Ipser}}},\ and\ \bibinfo {author} {\bibfnamefont {R.~D.}\ \bibnamefont
  {{Sorkin}}},\ }\bibfield  {title} {\bibinfo {title} {{Turning Point Method
  for Axisymmetric Stability of Rotating Relativistic Stars}},\ }\href
  {https://doi.org/10.1086/166043} {\bibfield  {journal} {\bibinfo  {journal}
  {\apj}\ }\textbf {\bibinfo {volume} {325}},\ \bibinfo {pages} {722} (\bibinfo
  {year} {1988})}\BibitemShut {NoStop}%
\bibitem [{Note4()}]{Note4}%
  \BibitemOpen
  \bibinfo {note} {In the original argument of \cite {1982ApJ...257..847S} a
  scalar $S$ to be maximized for an equilibrium state is related to variables
  $E^a$ as $dS = \beta _a dE^a$. As in \cite {1988ApJ...325..722F} we take $S =
  -M$ and an equilibrium minimize it. Then we have $E^a=(-N_B, -N_X)$ and
  $\beta _a=(\mu _B, \mu _X)$. Since an equilibrium star satisfies $dM = \mu _B
  dN_B + \mu _X dN_X$ we have $dM=\mu _X dN_X$ on the sequence with baryonic
  mass being fixed. Thus the stability condition reads $\protect \frac {d\mu
  _X}{d\lambda }\protect \frac {dM}{d\lambda }>0$}\BibitemShut {NoStop}%
\bibitem [{\citenamefont {Kouvaris}(2008)}]{PhysRevD.77.023006}%
  \BibitemOpen
  \bibfield  {author} {\bibinfo {author} {\bibfnamefont {C.}~\bibnamefont
  {Kouvaris}},\ }\bibfield  {title} {\bibinfo {title} {{WIMP} annihilation and
  cooling of neutron stars},\ }\href
  {https://doi.org/10.1103/PhysRevD.77.023006} {\bibfield  {journal} {\bibinfo
  {journal} {Phys. Rev. D}\ }\textbf {\bibinfo {volume} {77}},\ \bibinfo
  {pages} {023006} (\bibinfo {year} {2008})}\BibitemShut {NoStop}%
\bibitem [{\citenamefont {{Luo}}\ \emph {et~al.}(2008)\citenamefont {{Luo}},
  \citenamefont {{Bai}},\ and\ \citenamefont {{Zhao}}}]{2008ChPhL..25.2378L}%
  \BibitemOpen
  \bibfield  {author} {\bibinfo {author} {\bibfnamefont {X.-L.}\ \bibnamefont
  {{Luo}}}, \bibinfo {author} {\bibfnamefont {H.}~\bibnamefont {{Bai}}},\ and\
  \bibinfo {author} {\bibfnamefont {L.}~\bibnamefont {{Zhao}}},\ }\bibfield
  {title} {\bibinfo {title} {{GENERAL: Double Degenerate Stars}},\ }\href
  {https://doi.org/10.1088/0256-307X/25/7/013} {\bibfield  {journal} {\bibinfo
  {journal} {Chinese Physics Letters}\ }\textbf {\bibinfo {volume} {25}},\
  \bibinfo {pages} {2378} (\bibinfo {year} {2008})}\BibitemShut {NoStop}%
\end{thebibliography}%

\end{document}